\pdfoutput = 1

\documentclass[doublecol]{epl2}

\usepackage{color}
\definecolor{greencolor}{rgb}{0,0.5,0.2}
\definecolor{redcolor}{rgb}{.7,0.,0.}
\definecolor{bluecolor}{rgb}{0,0.,1.}
\definecolor{greycolor}{rgb}{.5,.5,.5}

\usepackage{hyperref}

\title{Unveiling the Relationship Between Complex Networks Metrics and Word Senses}
\shorttitle{Complex Networks for Word Sense Disambiguation} 

\author{Diego R. Amancio\inst{1} \and Osvaldo N. Oliveira Jr.\inst{1} \and Luciano da F. Costa\inst{1}}
\shortauthor{D. R. Amancio \etal}

\institute{
  \inst{1} Institute of Physics of S\~ao Carlos \\
	University of S\~ao Paulo, P. O. Box 369, Postal Code 13560-970 \\
	S\~ao Carlos, S\~ao Paulo, Brazil \\ \ \\ \\
}

\abstract{
The automatic disambiguation of word senses {(i.e., the identification of which of the meanings is used in a given context for a word that has multiple meanings)} is essential for such applications as machine translation and information retrieval, and represents a key step for developing the so-called Semantic Web. Humans disambiguate words in a straightforward fashion, but this does not apply to computers. In this paper we address the problem of Word Sense Disambiguation (WSD) by treating texts as complex networks, and show that word senses can be distinguished upon characterizing the local structure around ambiguous words. Our goal was not to obtain the best possible disambiguation system, but we nevertheless found that in half of the cases our approach outperforms traditional shallow methods. We show that the hierarchical connectivity and clustering of words are usually the most relevant features for WSD. The results reported here shine light on the {relationship} between semantic and structural parameters of complex networks. {They also indicate that when combined with traditional techniques the complex network approach may be useful to enhance the discrimination of senses in large texts}.
}

\begin{document}

\maketitle

\section{Introduction}

    {Many statistical methods are now used to investigate language~\cite{cs} in attempts to understand empirical findings such as the Zipf's Law~\cite{manning}, and model syntactic and semantic relationships between words or passages ~\cite{apl,cancho,summ,amancio,amancio2,mihalcea,kenett,sigman,steyvers,lacalle,consistencia}. The numerous studies encompass complex networks~\cite{cn1} (CN) representing texts in several applications, including summarization~\cite{summ}, assessment of quality of machine translators~\cite{amancio}, authorship recognition~\cite{amancio2}, keyword extraction~\cite{mihalcea}, topic identification~\cite{mihalcea2} and segmentation~\cite{mihalcea3}. In this paper we assess the use of complex network concepts for Word Sense Disambiguation (WSD), which is a crucial task for the Semantic Web~\cite{semanticweb} and for machine translation~\cite{weaver}. We analyze $10$ ambiguous words with $16$ topological measurements and show a strong relationship between senses and local features of complex networks. Indeed, for some of the ambiguous words the distinguishability with the CN approach is better than that obtained with the traditional analysis of neighbors. From an analysis of feature relevance, we found that the strength of connection of neighbors in higher hierarchies and the hierarchical clustering coefficient are the most efficient metrics to discriminate word senses.}

    \section{Typical Approaches to Word Sense Disambiguation}

    The WSD problem has been widely studied~\cite{review} by computer scientists and researchers interested in Natural Language Processing~\cite{manning} tasks. Even though humans can readily discriminate specific senses of a word, this is not the case of a computer. In fact, WSD is considered as one of the most complex problems in Artificial Intelligence ~\cite{mallery1988}. The two conventional approaches to WSD are: i) the deep paradigm based on a large amount of linguist knowledge (e.g. dictionaries, thesaurus or semantic networks); and ii) the shallow paradigm which makes use of statistical techniques. The deep paradigm is in theory the best strategy as it mimics human thinking, but in practice methods requiring knowledge bases do not achieve the best performance because there is still no database that can cover the human knowledge. Moreover, this paradigm is often impracticable because the manual creation of knowledge bases is an expensive, time consuming endeavor. In contrast, simpler methods such as those based on the analysis of contexts surrounding ambiguous words~\cite{yarowski1993} have led to better performance. One of the most popular algorithms, referred to as Lesk~\cite{lesk}, assumes that words in a given neighborhood tend to share a common topic, an assumption that is used by other algorithms~\cite{review}. Actually, the analysis of contexts based on the recurrence of nearby components is so efficient that it has even been employed to decipher encoded manuscripts~\cite{copiale}.

    Networks have also been applied to the WSD task, and some of the network-based algorithms are now close to the state-of-the-art in disambiguation. One of the earliest works date back to $1968$ and uses the network structure to store knowledge in the form of a semantic memory~\cite{gbased}. Other examples include the application of random walks~\cite{randomWalk} in semantic networks whose nodes are linked according to semantic relations provided by WordNet\cite{mihalcea2004}. With a different approach, the HyperLex algorithm~\cite{veronis2004} connect words that co-occur in a given paragraph and use the weight of edges (given by the relative frequency of occurrence of the corresponding connected nodes) to disambiguate words. Although these algorithms use the network representation in processing steps, they differ substantially from our strategy because they all consider the label of nodes while we focus on the characterization of local structural properties.

    \section{Methodology}

    \subsection{Database}

    In the experiments, we used a set of $18$ books to retrieve $10$ ambiguous words (save, note, march, present, jam, ring, just, bear, rock and close), which were manually disambiguated. {The only criterion in choosing these words was the quite distinct meanings of each word, which minimizes possible inaccuracies in the manual disambiguation}. The list of word senses and books are given respectively in Tables S1 and S2 of the \href{http://dl.dropbox.com/u/2740286/epl_SI.pdf}{Supplementary Information} (SI). {The text in the books was represented as networks, as explained below}.

    \subsection{Modeling Texts as Complex Networks}

    The model used to represent text is known in the literature as co-occurrence or adjacency networks~\cite{amancio,amancio2}. Basically, words are represented as nodes, which are directionally linked according to the natural reading order. In other words, if a word $i$ appears immediately before word $j$ in the text, then there will be the $i \rightarrow j$ edge in the network. When a given association $i \rightarrow j$ is repeated in the text, the weight of the corresponding edge is incremented. Before creating nodes and edges, stopwords (prepositions, articles and other high-frequency words with little semantic meaning) are removed (the full list of disregarded stopwords is shown in the \href{http://dl.dropbox.com/u/2740286/epl_SI.pdf}{SI}). In addition, the remaining words are converted to their canonical form in order to group words with different inflections referring to a same concept.



    Mathematically, the text network is defined by the matrix $W = \{ w_{ij} \}$, whose element $w_{ij}$ counts the number of times the word $i$ appeared before the word $j$. When defining some of the complex networks measurements we also employed the non-oriented, non-weighted version, represented by the matrix $A = \{a_{ij}\}$ so that $a_{ij} = 1$ if $i$ appeared at least once as a neighbor of $j$ (regardless of the position) and $a_{ij} = 0$ otherwise. When a word repeatedly appears in the same text, it is considered as the same node in the corresponding network. But this procedure is not adopted for the ambiguous words under analysis, i.e. each occurrence is taken as a distinct node in the network, so that it is possible to characterize each occurrence of an ambiguous word to correlate its structural features with its meanings.

\subsection{Characterization of Senses Through Complex Networks Features} \label{cnmeasurements}

    To characterize the local structure of an ambiguous word, we used a set of $16$ complex network local measurements~\cite{cn1}. The simplest measurement is the degree $k_1$, i.e., the number of connections (without considering the weight of the edges). In terms of the adjacency matrix $A$, the degree is computed as
    \begin{equation}
        k_1(i) = \sum_{j} a_{ij}.
    \end{equation}
    The weighted version of $k_1$, which considers the strength~\cite{cn1} of the links, is given by
    \begin{equation}
        s_1(i) =  \sum_{j} w_{ij}.
    \end{equation}

    Extensions of these two measurements were considered through the analysis of further hierarchies~\cite{concentrical} for the hierarchical expansion usually yields better network characterization~\cite{amancio,concentrical}. The expansion of a given node is made by merging the node under analysis with its neighbors in a single node, keeping the external connections of the neighbors~\cite{amancio,concentrical}. This procedure is then repeated to generate deeper expansions. This hierarchical characterization was adopted for both $k_1$ and $s_1$, where the $m$-th expansion is represented as $k_{m+1}$ and $s_{m+1}$. We have not made explicit use of $k_{m+1}$ and $s_{m+1}$ when $m=0$ because these measurements take constant values as a consequence of considering each occurrence of an ambiguous word as a single node.

    In addition to the local measurements, we quantified the connectivity of nodes to their neighbors. Analogously to the adoption of further hierarchies, the study of topological properties of neighbors also yields better network characterization~\cite{eplcosta}. Indeed, neighbors have played a key role in many algorithms, such as the PageRank~\cite{pagerank} algorithm and its variations. In this paper, the following neighborhood-based measurements were employed: the average degree and strength of the neighbors ($\langle k_n \rangle$ and $\langle s_n \rangle$, respectively) and their standard deviations ($\Delta k_n$ and $\Delta s_n$). Another structural measurement used was the clustering coefficient ($C$), which is proportional to the fraction of triangles over the total number of connected triads. More specifically, the clustering is computed as:
    \begin{equation} \label{aglomeracao}
    	C_1(i) =  \frac{ 3~\sum_{k > j > i} a_{ij} a_{ik} a_{jk} }{ \sum_{k > j > i} a_{ij} a_{ik} + a_{ji} a_{jk} + a_{ki} a_{kj} }.
    \end{equation}
    It is known that a correlation exists between the number of semantic contexts where a word appears and its clustering coefficient~\cite{amancio2}. Since word senses might be related to the number of contexts (because distinct senses could appear in different contexts) this measurement may be useful in the disambiguation task. Similarly to the degree and strength measurements, we expanded the hierarchies up to $m = 3$ to compute this measurement.

    The local structure was also examined with shortest paths (or geodesic paths) between two nodes, which are paths whose sum of the edge weights is minimum. If $d_{ij}$ is the shortest path between nodes $v_i$ and $v_j$ in the adjacency matrix $A$, then the average shortest $l$ path length for $v_i$ is:
    \begin{equation}
    \label{geodesico}
	   l(i) = \frac{1}{n-1} \sum_{j=1}^{n} d_{ij}.
    \end{equation}
    In networks of text, the shortest path quantifies the centrality of a word according to its distance to the most frequent words~\cite{amancio2}. We chose to use this measurement to verify if the distance from an ambiguous word to the core-content concepts of the books can be used do distinguish senses. Shortest paths were also employed to compute the betweenness of words ($B$). Let $\eta_{st}^i$ be the number of shortest paths between nodes $v_s$ and $v_t$ that pass through node $v_i$. If $g_{st}$ is the number of shortest paths between nodes $v_s$ and $v_t$, then $B$ is defined as:
    \begin{equation}\label{eq.bi}
	   B(i) = \sum_{s} \sum_{t} \frac{\eta_{st}^i}{g_{st}}.
    \end{equation}
    Even though we are aware of the correlation between $B$ and $k$ ($B \sim k^{\eta}$)~\cite{amancio2,dynamical} in large networks, the possible distinct values of $B$ taken for ambiguous words will not reflect differences in word frequency, because $k=2$ for each occurrence of an ambiguous word. Actually, $B$ will reflect the ability of words to connect different network regions~\cite{dynamical,goh} or different contexts~\cite{amancio2}.

    The $16$ measurements employed to characterize the local structure of ambiguous words are summarized in Table \ref{tab.summ}, which contains the measurements classified according to their type (connectivity, clustering, neighborhood and paths), their notation and complexity.

    \begin{table}[h]
    \centering
    \caption{\label{tab.summ} {List of measurements employed to characterize the local structure of nodes representing ambiguous words in complex networks. The last column indicates the time complexity, i.e., the time taken to compute each measurement as a function of the number of nodes $n$ and edges $e$. For the clustering coefficient, the time complexity ranges between O($n$) and O($n^2$) because it depends explicitly both on $\langle k \rangle$ and $\langle k^2 \rangle$. Further details regarding measurements and time complexity can be found in Ref.~\cite{newman}}.}
        \begin{tabular}{|c|c|c|}
            \hline
            \textbf{Group} & \textbf{Notation} &  \textbf{Complexity} \\
            \hline
            Connectivity   & $k_2$, $k_3$ and $k_4$ & O($n$ + $e$) \\
                           & $s_2$, $s_3$ and $s_4$ & O($n$ + $e$) \\
            \hline
            Clustering     & $C_1$, $C_2$, $C_3$ and $C_4$ & [O($n$),O($n^2$)] \\
            \hline
            Neighborhood   & $\langle k_n \rangle$ and $\Delta k_n$ & O($n^2$) \\
                           & $\langle s_n \rangle$ and $\Delta s_n$ & O($n^2$) \\
            \hline
            Paths          & $l$ & O($n^2$) \\
                           & $B$ & O($n^2$) \\
            \hline
        \end{tabular}
    \end{table}

    To verify if the description provided by the measurements above is useful for the WSD task, we used machine learning algorithms that induce classifiers from the training set provided for each word. The quality of the results was then evaluated using the 10-fold cross validation technique~\cite{validation}, which was chosen because it is robust in the sense that the training set is always different from the evaluation set. Thus, it prevents that overfitted inductors take high values of accuracy rate. $3$ inductor algorithms were used: the C4.5 algorithm~\cite{bishop}, which generates trees based on the gain provided by each feature; the Naive Bayes algorithm~\cite{bishop}, which uses the Bayes theorem; and the k nearest neighbor algorithm~\cite{bishop} (kNN), which classifies an external unknown instance according to the most similar instance of the training database in a normalized space including all features. Details regarding algorithms and the cross validation technique are given in the \href{http://dl.dropbox.com/u/2740286/epl_SI.pdf}{SI}.

\section{Results and Discussion}

\begin{figure}
        \begin{center}
            \onefigure[width=0.45\textwidth]{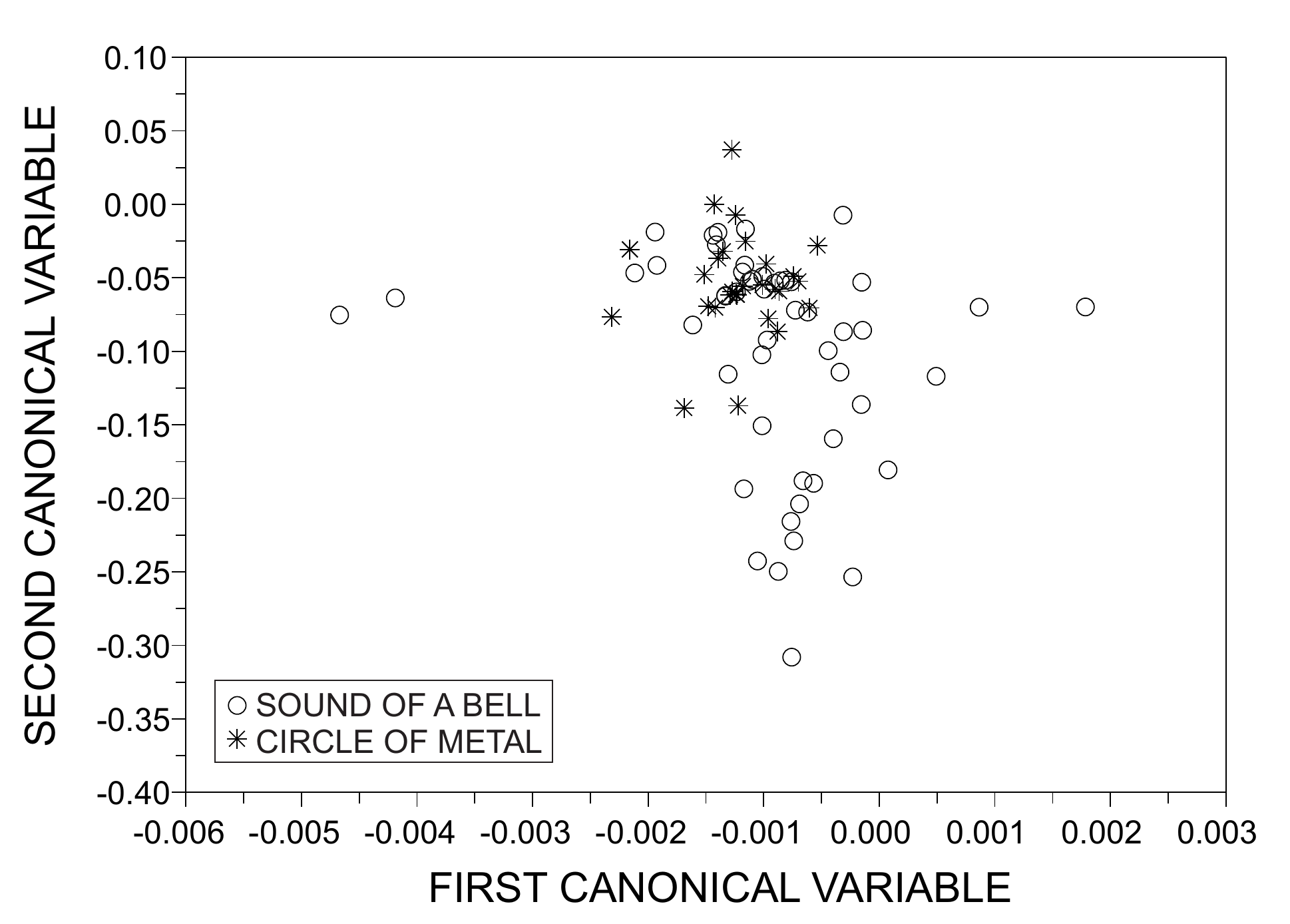}
        \end{center}
        \caption{\label{fig2} Canonical Analysis Projection for the word ``ring''. The senses considered were i) ring of a bell and ii) circle of metal. While the use of the sense `circle of metal' is more regular as revealed by the low dispersion, the sense `sound of a bell' tends to be more heterogeneous.}
        \end{figure}

        The $10$ ambiguous words were characterized with complex networks measurements to verify if senses can be inferred from a topological analysis. Table \ref{tab.1} shows in the second and third columns, respectively, the accuracy rate and the corresponding p-value $\alpha_{cn}$ relative to a classification performed by assigning the most common (i.e. the most frequent) sense to the ambiguous word. A significant accuracy ($\alpha_{cn} < 5.0~10^{-2}$) could be observed in $9$ out of the $10$ words. An example of scatter plot depicting the discrimination obtained for the word ``ring'' is shown in Figure~\ref{fig2}, where each axis represents a linear combination of the $16$ measurements provided by the Canonical Variable Analysis technique~\cite{duda}. These results confirm the relationship between local characteristics of adjacency networks and word senses, reinforcing the suitability of complex network methods to relate structure and semantics. {We believe that the ability to distinguish senses is at least partially due to the fact that co-occurrence networks probably imply syntactic factors~\cite{gamalo} that are reflected on the semantic relations~\cite{gamalo2}. This relationship, however, is still difficult to establish because there is no consolidated interpretation for the measurements of word adjacency networks (see e.g. Ref.~\cite{amancio2})}.

        \begin{table*}
        \centering
        \caption{\label{tab.1} {Results from characterizing ambiguous words with structural complex networks measurements. The accuracy rate and the p-value, considering as null model a classifier based on the most common sense, and the best classifier are shown for all words. The minimum set of measurements considered in each classifier is shown in the last column. Apart from the word ``close'', all the classifications achieved significant accuracy rates ($\alpha_{cn} < 5.0~10^{-2}$). Interestingly, even though $16$ measurements were employed in the characterization, the top classifiers were induced with $5$ measurements or less.}}
            \begin{tabular}{|c|c|c|c|c|}
            \hline
            \textbf{Word} & \textbf{Acc. Rate} & \textbf{$\alpha_{cn}$} & \textbf{Best Ind.} & \textbf{Minimum Set of Measurements} \\
            \hline
            save            & 87.64 \%  &  $1.6~10^{-3}$ & kNN  & $\langle s_n \rangle$ and $l$ \\
            note            & 84.53 \%  &  $4.7~10^{-2}$ & kNN  & $C_3$, $s_3$ and $\Delta s_n$\\
            march           & 86.95 \%  &  $1.9~10^{-3}$ & kNN  & $C_3$, $\langle k_n \rangle$, $\sigma$\\
            present         & 71.14 \%  &  $2.2~10^{-2}$ & kNN  & $s_4$, $\langle k_n \rangle$, $\langle k_n \rangle$ and $\Delta s_n$ \\
            jam             & 100.0 \%  &  $6.0~10^{-3}$ & kNN  & $C_4$, $s_4$ and $\Delta k_n$ \\
            ring            & 84.61 \%  &  $2.8~10^{-4}$ & kNN  & $C_4$, $k_4$, $s_4$, $\Delta k_n$ and $l$ \\
            just            & 51.28 \%  &  $1.6~10^{-2}$ & kNN  & $C_2$ and $C_3$ \\
            bear            & 61.95 \%  &  $8.0~10^{-3}$ & kNN  & $C_4$, $k_3$ and $s_4$ \\
            rock            & 79.30 \%  &  $9.7~10^{-9}$ & C4.5 & $k_2$, $s_2$, $k_3$, $k_4$ and $s_4$ \\
            close           & 72.20 \%  &  $8.0~10^{-2}$ & kNN  & $C_2$, $s_2$, $\langle s_n \rangle$ and $\Delta s_n$ \\
            \hline
            \end{tabular}
        \end{table*}


        Although our primary goal was not a search for the best possible disambiguation system, we compare our results with the traditional approach based on the analysis of frequency of nearby words. Classifiers were induced using attributes that represent the frequency of the $5$, $20$ or $50$ words surrounding the ambiguous word. The lowest p-values and the top classifiers are shown in Table \ref{tab.2}. For the first $5$ words the CN approach outperformed the traditional method. This means that the local structure can be even more relevant than the frequency analysis of neighbors. Obviously, we are not suggesting the CN approach to replace the approaches based on semantic information provided by neighbors, since complex network measurements are statistically reliable only when computed in large texts. Still, it could be valuable to combine both strategies in disambiguation systems. As for the methods and algorithms, the differences regarding the best classifier are worth noting. The CN approach performs better with the kNN algorithm, while the Naïve Bayes algorithm is better in the traditional approach. These differences occur probably because of the distinct number of attributes in each approach. Finally, in {6 out of the 10 words considered, the traditional analysis with $5$ neighbors outperformed the classifiers with larger numbers of neighbors.}

        \begin{table}[h]
        \centering
        \caption{\label{tab.2} {Results from characterizing ambiguous words with the traditional approach (second column) and with the CN approach (third column). The p-value ($\alpha_{tr}$ for the traditional approach and $\alpha_{cn}$ for the CN-based method) was computed considering as null model a classifier based on the most common sense. The best classifier algorithm in the fourth column refers to the traditional approach.}}
            \begin{tabular}{|c|c|c|c|}
            \hline
            \textbf{Word}   & \textbf{$\alpha_{tr}$} & \textbf{$\alpha_{cn}$} & \textbf{Best Ind.}  \\
            \hline
            save            & $6.2~10^{-1}$     & $1.6~10^{-3}$ & kNN \\
            note            & $3.8~10^{-1}$     & $4.7~10^{-2}$ & kNN \\
            march           & $1.4~10^{-2}$     & $1.9~10^{-3}$ & kNN \\
            present         & $6.8~10^{-2}$     & $2.2~10^{-2}$ & Naive Bayes \\
            jam             & $1.0~10^{-2}$     & $6.0~10^{-3}$ & Naive Bayes \\
            ring            & $3.8~10^{-9}$     & $2.8~10^{-4}$ & Naive Bayes \\
            just            & $1.8~10^{-4}$     & $1.6~10^{-2}$ & Naive Bayes \\
            bear            & $3.3~10^{-5}$     & $8.0~10^{-3}$ & C4.5        \\
            rock            & $< 1.0~10^{-10} $ & $9.7~10^{-9}$ & Naive Bayes \\
            close           & $< 1.0~10^{-10} $ & $8.0~10^{-2}$ & Naive Bayes \\
            \hline
            \end{tabular}
        \end{table}

\begin{figure}
    \begin{center}
        \onefigure[width=0.45\textwidth]{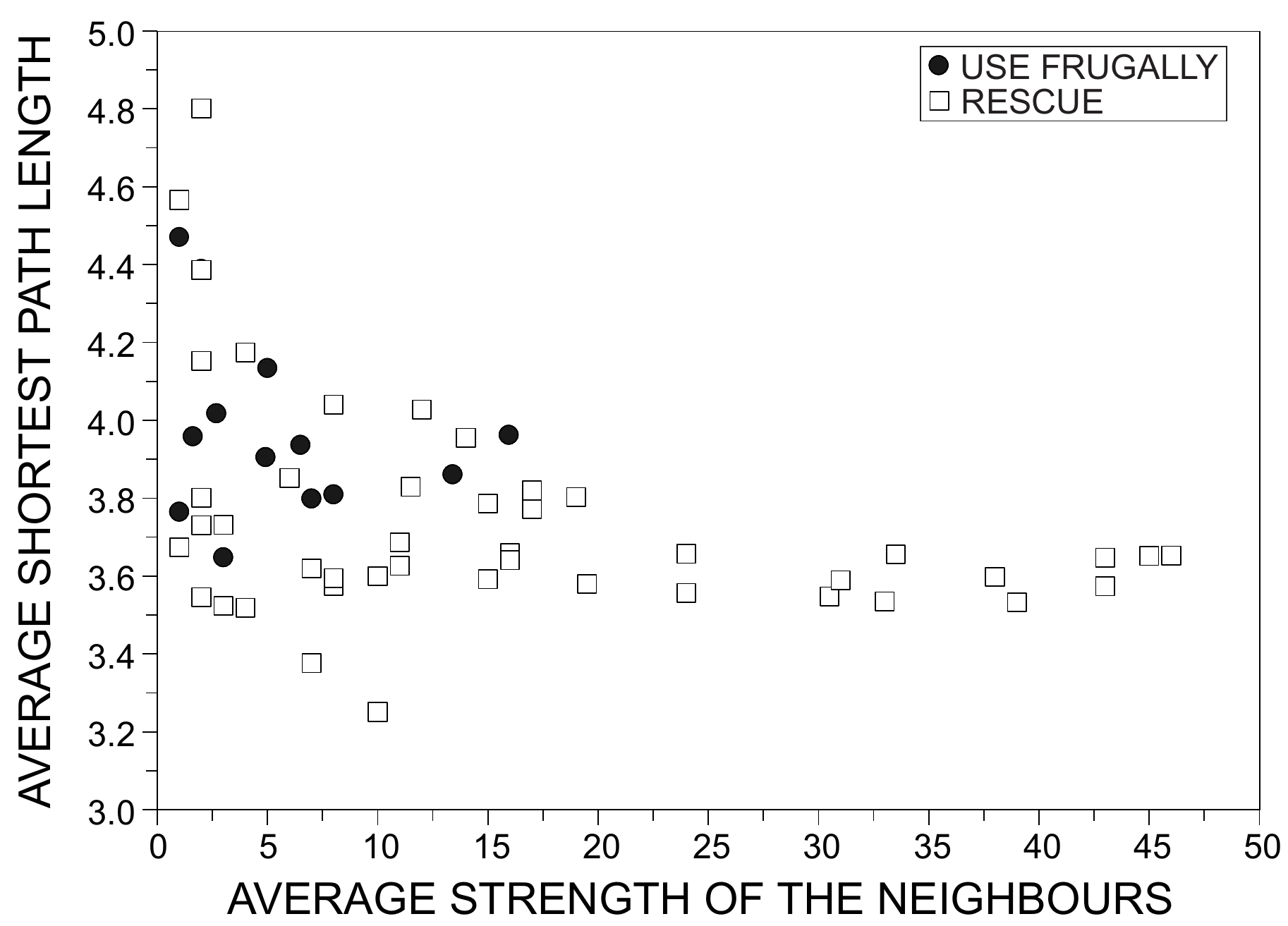}
    \end{center}
    \caption{\label{fig.3}  Scatter plot for the word ``save''.
    While one sense is characterized by high $l$ and low $\langle s_n \rangle$; the other sense usually takes low values of $l$.}
    \end{figure}

    The contribution from each network metric in discriminating word senses was estimated by first finding the smallest subset of measurements generating the best classifiers (see last column of Table \ref{tab.1}). Although we used $16$ measurements to characterize nodes, the best accuracy rates were obtained with a maximum of $5$ measurements. Strikingly, in some cases only two measurements were already sufficient to provide a reasonable distinction, as indicated in the scatter plot for the word ``save'' in Figure \ref{fig.3}). {Quantitatively, the relevance of each metric (i.e. feature) for disambiguating the words was calculated in two ways: using the Kullback-Leibler (KL) divergence and the method based on the Mann Whitney U (MWU) test~\cite{mann}}. {While in the latter features are evaluated individually, in the former the interaction between features is considered. Thus, it is possible to identify cases where features are useful only when combined with others. Details regarding the KL divergence and the MWU test are given in the \href{http://dl.dropbox.com/u/2740286/epl_SI.pdf}{SI} and in Ref.~\cite{amancio2}}.

The rankings shown in Table \ref{tab.4} indicate that the relevance of a metric varies from word to word. In addition, a metric may be relevant when analyzed individually but not so if combined with other attributes because some features included in the first method may not be included in the second one. According to the KL divergence, the most frequent relevant  metrics (i.e. the ones which appear among the top $3$) are $C_3$ and $l$. For the MWU test, the clustering computed at high hierarchical level ($C_4$) is also a relevant feature along with $s_3$ and $s_4$. Therefore, these results suggest that meanings are often correlated with the strength (or frequency) of higher-order neighbors and with the degree of interconnection of neighbors. Interestingly, $l$ was not so relevant when combined with other features as it appeared only a few times in the MWU ranking.

    \begin{table*}
    \centering
    \caption{\label{tab.4} Rankings of metrics obtained with the Kullback-Leibler method and with the Mann-Whitney U test.}
    \begin{tabular}{|c|c|c|c||c|c|c|c|c|}
    \hline
    Method   & \multicolumn{3}{|c||}{Kullback-Leibler} & \multicolumn{5}{|c|}{Mann-Whitney U test}\\
    \hline
    Word     & \textbf{$1^{st}$} & \textbf{$2^{nd}$} & \textbf{$3^{rd}$} & \textbf{$1^{st}$} & \textbf{$2^{st}$} & \textbf{$3^{rd}$} & \textbf{$4^{th}$} & \textbf{$5^{th}$} \\
    \hline
    bear &$C_2$ & $ \langle s_n \rangle$ & $\langle k_n \rangle$ &   $\sigma$  &  $s_3$  &  $k_2$ & $k_4$  & $\Delta k_n$ \\
    jam     & $C_4$ & $C_3$ &   $ \langle s_n \rangle$ & $s_3$     &  $C_4$  &  $C_2$    &   $s_4$  & $k_4$ \\
    just    & $C_4$ & $C_3$ & -- & $k_3$     &  $s_4$  &  $C_4$    &   $s_3$  & $k_4$        \\
    march   & -- & -- & -- & $\sigma$  &  $C_4$  &  $k_2$    &   $s_4$  & $k_4$        \\
    ring    & $l$ & -- & -- & $\sigma$  &  $C_4$  &  $k_4$    &   $s_3$  & $l$          \\
    present & -- & -- & -- & $\langle s_n \rangle$ & $C_4$ & $\langle k_n \rangle$ & $s_4$ & $k_3$ \\
    close   & -- & -- & -- & $C_3$   &   $l$ &   $s_4$ &    $\langle k_n \rangle$ & $k_4$ \\
    note    & $C_3$ & $s_4$ & $s_2$ & $s_3$   &   $\sigma$ & $\Delta k_n$ & $\Delta s_n$ & $s_2$ \\
    save    & $l$ & -- & -- & $\Delta s_n$ & $s_2$ & $l$ & $k_3$ & $C_4$ \\
    rock    & $l$ & $s_2$ & $k_2$ & $\langle k_n \rangle$ & $k_2$ & $s_2$ & $s_4$ & $s_3$ \\
    \hline
    \end{tabular}
    \end{table*}


    \section{Conclusion}

    In this paper we have verified the suitability of the complex network model for the word sense disambiguation task in large texts. Upon characterizing the local structure of nodes representing ambiguous words, we obtained significant discrimination, which means that different senses affect the structural organization of complex networks. {Strikingly,} the discrimination was so effective for some words that the topological characterization outperformed traditional shallow methods. In general, the hierarchical characterization of the clustering and connectivity measurements were the most relevant features for WSD, even though the ranking of metrics varied from word to word. The analysis here may shed light on the relationship between structure of complex networks and semantics. {From a practical standpoint, the methodology described might be useful in hybrid approaches to improve state-of-the-art disambiguating systems. Given an extensive set of texts, it is possible to obtain networks with the local characterization of nodes representing words whose meaning is known beforehand. Then, an ambiguous word of a book could be disambiguated by assigning meanings according to the semantic (traditional approach) and topological features (CN approach) provided by the training set.} {In future works, we plan to use wider window sizes to connect words and additional complex networks measurements~\cite{cn1}, such as weighted versions of the shortest path, clustering coefficient, and betweenness along with CN-based classification algorithms~\cite{thiago} to improve the performance of disambiguation systems in long texts. Also, we shall study the influence on the results when the proposed methodology is applied to other languages.}

    %

\acknowledgments
The authors acknowledge the financial support from CNPq (Brazil) and FAPESP (2010/00927-9).

\end{document}